\documentclass[twocolumn,showpacs,preprintnumbers]{revtex4}
\usepackage{graphicx}
\usepackage{dcolumn}
\usepackage{bm}
 
\begin{document}
 
\preprint{}
\title{``Quasiparticle Charge" in Superconductors: Effect of 
Mott Physics or Hidden Order Parameter ?}
\author{Hae-Young Kee and Yong Baek Kim}
\affiliation{Department of Physics, University of Toronto, Toronto, Ontario, 
Canada M5S 1A7}
\date{\today}

\begin{abstract}
The renormalization factor,
dubbed ``quasiparticle charge'' $Z_e$, of the coupling between the supercurrent and 
quasiparticles  was analyzed in the context of high temperature cuprates
in a  recent paper by  Ioffe and Millis.\cite{millis}
They observed that $Z_e$ in cuprates deviates from the BCS value ($Z_e=1$),
which was interpreted as the proximity effect near a Mott insulator. 
Here we show that the deviation from $Z_e=1$ can occur, in general, even in
the absence of quasiparticle interactions, when the superconducting order 
coexists with another order parameter with the same internal symmetry.
As an example, we compute the coefficient of the linear temperature dependence 
of the superfluid density when the d-wave superconducting state coexists with
the orbital antiferromagnetic state (d-density wave), and find that $Z_e$ varies 
from $1/\sqrt{2}$ to 1.
\end{abstract}
 
\maketitle

\def\be{\begin{equation}}
\def\ee{\end{equation}}
\def\bea{\begin{eqnarray}}
\def\eea{\end{eqnarray}}

\section{Introduction}
The behavior of charge transport in both the normal and superconducting phases of 
cuprates is one of the puzzling issues in high temperature superconductors.
In a recent paper, Ioffe and Millis discussed the effect of proximity
to a Mott insulator on charge transport in the superconducting state 
of cuprates. \cite{millis} They pointed out that the renormalization factor, 
dubbed ``quasiparticle charge'' $Z_e$, of the the coupling between the 
supercurrent and quasiparticles is renormalized by quasiparticle interactions
and contains useful information about strong correlations near a Mott insulator.
\cite{larkin64,leggett65}

Even though the real quasiparticle charge is not conserved in superconductors and $Z_e$
is not the conventionally defined charge, we will still use ``quasiparticle charge'' to 
call the renormalization factor mentioned above for convenience. 
As pointed out in the past \cite{millis,larkin64,leggett65}, $Z_e$ appears in the linear 
temperature dependence of the superfluid density in d-wave superconductors (dSC). 
The low temperature superfluid density of d-wave superconductors is given by
\begin{equation}
\rho_s (T) = \rho_s (0) - \alpha T,
\label{sfd}
\end{equation}
where
\begin{equation}
\alpha = \frac{ (\ln 2) Z_e^2 v_F}{2 \pi v_{\Delta}}.
\label{alpha1}
\end{equation}
Here $v_F$ is the Fermi velocity and $v_{\Delta}$ the gap velocity
defined as $v_{\Delta} = \Delta_0 a/\sqrt{2}$, where $\Delta_0$ is
the maximum value of superconducting order parameter.

The behavior of the superfluid density in cuprates has been an important
subject of debate. Lee and Wen observed that, in cuprates, $\alpha$ does 
not strongly depend on the doping concentration, $x$, while the zero temperature
superfluid density is proportional to $x$.\cite{lee97,wen}
This would be consistent with the phenomenology that the quasiparticles in 
the superconducting state of cuprates still couples to external 
electromagnetic field in 
the usual BCS manner even in the presence of strong correlation near 
a Mott insulator.\cite{lee97}
On the other hand, two different strong coupling approaches to the t-J model 
lead to different behaviors of the superfluid density. 
The U(1) mean field theory predicts $\alpha \propto x^2$ 
as well as $\rho_s(0) \propto x$ \cite{kotliar88}, implying that the 
``quasiparticle'' charge goes to zero as the Mott insulator is approached
while the SU(2) gauge theory lead to $\alpha \sim {\cal O}(1)$ with
the same $\rho_s(0)$.\cite{lee97,wen} 

Later on, Millis et al \cite{millis98} pointed out that $\alpha$ is in general
renormalized by quasiparticle interactions even in the conventional d-wave
superconductors. Millis et al further argued that the deviation from the standard
BCS value $Z_e=1$ should be regarded as an evidence of the strong correlation
present near a Mott insulator.\cite{millis98} 
Recently, Ioffe and Millis extracted the values of $v_F$ and $v_{\Delta}$ from 
the angle resolved photoemission spectroscopy (ARPES) and the specific heat data
of cuprates, and obtained the value of $Z_e$ for several samples of cuprates.
They found that $Z_e$ varies from $0.6$ to $1$ in cuprates, 
and argued that the deviation from the BCS 
value ($Z_e=1$) is due to strong quasiparticle interactions 
near a Mott insulator.

In this paper, we show that the deviation from the BCS value of $Z_e$
can also occur within the weak coupling approach without taking into account
the quasiparticle interactions,
when a superconducting state coexists with another
ordered state with the same internal symmetry.
In this case, the value of $Z_e$ contains informations about 
the additional order parameter.
As an example, we consider a system where
the d-wave superconducting order coexists with
the commensurate orbital antiferromagnetic order (d-density wave).\cite{relevance}
The quasiparticles in this state have the Dirac spectrum near
the nodes as the case of dSC and the superfluid density has the linear 
temperature dependence.
By computing the coefficient of the linear temperature
dependence of the superfluid density, we find that $Z_e$ in Eq.\ref{alpha1}
varies from $1/\sqrt{2}$ to $1$.
The deviation from the BCS
value ($Z_e=1$) in this case is {\it not} due to the quasiparticle interaction,
but due to the existence of the additional order parameter.
The superfluid density of the dSC coexisting with the dDW order was previously 
studied \cite{sumanta,han}, but the question of the renormalization factor $Z_e$ 
was not addressed. 
Although we studied a specific system, our results can be applied to
more general situation where a superconducting order coexists with
another order parameter with the same internal symmetry.

In the next section, we compute the current-current correlation 
function in the coexisting dSC and dDW states.
Those who are not interested in the details can skip the next section and 
jump to section III. In the third section, we provide the analytical 
expression of $\alpha$ and extract the value of $Z_e$. 
We compare our results with the consequences of other existing theories 
and discuss further implications in the final section. 

\section{Superfluid density in the coexisting dDW and dSC states}

In the phase where the dSC and dDW coexist, the quasiparticles keep
the Dirac spectrum as the case of dSC. In particular, when the chemical 
potential $\mu$ is zero or $\mu < T \ll {\rm min}(\Delta_0, W_0)$, 
the low energy nodal spectrum 
is well described by changing the gap velocity $v_{\Delta}$ to ${\tilde v}_{\Delta} 
=\sqrt{v_{\Delta}^2+v_W^2}$ where $v_W = W_0 a/\sqrt{2}$ and $W_0$ is the 
maximum value of the dDW order parameter. 
In this case, the gap velocity measured in ARPES will 
correspond to ${\tilde v}_{\Delta}$. The density of states will be also
modified, but in a trivial fashion; $v_{\Delta}$ is replaced by 
${\tilde v}_{\Delta}$. Thus one may expect that the coefficient of the
linear temperature dependence of the superfluid density is simply 
given by $\alpha = (\ln 2) v_F / (2 \pi {\tilde v}_{\Delta})$. If this 
is the case, we will not be able to extract an independent information
about $v_W$ and practically $Z_e=1$; experiments will not be able to
tell the difference between the simple dSC state and the coexisting
dSC/dDW states. We show below that this is not the case. 
$Z_e$ has an independent information about $v_W$ and in general 
it is not unity. The deviation from $Z_e=1$ is in fact due to the
existence of the dDW order parameter.  

We derive the result described above by directly computing the current-current
correlation function in a mean-field theory. 
We start from the mean-field Hamiltonian given by
\bea
H &= & \sum_{{\bf k},\sigma} (\epsilon_{\bf k}-\mu)
 c^{\dagger}_{{\bf k},\sigma} c_{{\bf k},\sigma}
+ \sum_{{\bf k},\sigma} i W_{\bf k}
 c^{\dagger}_{{\bf k},\sigma} c_{{\bf k}+{\bf Q},\sigma}
\cr
& & \hspace{0.5cm}
+ \sum_{{\bf k},\sigma}  \Delta_{\bf k}
 c^{\dagger}_{{\bf k},\sigma} c^{\dagger}_{-{\bf k},-\sigma}
+h.c.,
\eea
where
\bea
\epsilon_{\bf k} =  -{t \over 2} \left(\cos{k_x}+\cos{k_y} \right), \\
\Delta_{\bf k} =  {\Delta_0 \over 2} \left(\cos{k_x}-\cos{k_y} \right), \\
W_{\bf k} = {W_0 \over 2} \left(\cos{k_x}-\cos{k_y} \right).
\eea
Here $\Delta_{\bf k}$ and $i W_{\bf k}$ are the order parameters of
the dSC and dDW respectively.
Using the Nambu basis defined as follows,
\be
\Psi^{\dagger}_{\bf k}=( c^{\dagger}_{{\bf k}\uparrow},
 c^{\dagger}_{{\bf k}+{\bf Q}\uparrow},
 c_{-{\bf k}\downarrow},
 c_{-{\bf k}-{\bf Q}\downarrow}),
\ee
the current operator can be written as
\bea
j_{\alpha} ({\bf q}) &= &  t \sum_{\bf k}
\sin{(k_{\alpha}+q_{\alpha}/2)}
\Psi^{\dagger}_{\bf k} (\rho_3 I) \Psi_{{\bf k}+{\bf q}}
\cr
 &-&  i W_0 \sum_{\bf k} \sin{(k_{\alpha}+q_{\alpha}/2)}
\Psi^{\dagger}_{\bf k} (i \rho_2 \sigma_3) \Psi_{{\bf k}+{\bf q}},
\eea
where $\alpha = x, y$. Here $\rho$'s and $\sigma$'s are the 
Pauli matrices and $\rho_i \sigma_j \equiv \rho_i \otimes \sigma_j$.

The superfluid density can be computed from the current-current 
correlation function.\cite{schrieffer}
\be
\rho_s(T) = \langle -K_x \rangle - 
\lim_{q \rightarrow 0} \lim_{\nu \rightarrow 0}
\Lambda_{xx} ({\bf q},\nu),
\label{super}
\ee
where $\langle -K_x \rangle$ is the diamagnetic contribution
and $\Lambda_{xx} ({\bf q},\nu)$ is the paramagnetic current-current
correlation function. 
The diamagnetic part can be computed as
\bea
\langle -K_x \rangle &=& \sum_{\bf k} 
\frac{\epsilon_k+ W_k}{\Omega_k} \sin{k_x}
\cr
&& \hspace{-0.5cm} \times
\left( \frac{\Omega_k-\mu}{E_{1k}}\tanh{\frac{E_{1k}}{2T}}
+\frac{\Omega_k + \mu}{E_{2k}} \tanh{\frac{E_{2k}}{2T}}
\right),
\eea
where
\bea
E_{1k,2k} &=& \sqrt{(\Omega_k \mp \mu)^2 + \Delta_k^2},
\cr
\Omega_k &=& \sqrt{\epsilon_k^2+W_k^2}.
\eea

\begin{widetext}
The paramagnetic contribution contains the information about
$Z_e$ and can be obtained from
\bea
\Lambda_{xx}({\bf q}, i\nu_m) 
&=&
t^2 T \sum_n \sum_{\bf k}
\sin{k_x} \sin{(k_x+q_x)} 
{\rm Tr} [ G({\bf k},i \omega_n) (\rho_3 I)
G({\bf k}+{\bf q},i\omega_n+i\nu_m) (\rho_3 I)]
\cr
&&+ W_0^2 T \sum_n \sum_{\bf k}
\sin{k_x} \sin{(k_x+q_x)}
{\rm Tr} [ G({\bf k},i \omega_n) (\rho_2 \sigma_3)
G({\bf k}+{\bf q},i\omega_n+i\nu_m)
 (\rho_2 \sigma_3) ],
\label{corr}
\eea
where
\be
G^{-1}({\bf k},i\omega_n)= 
i \omega_n I +\epsilon_{\bf k} \rho_3 \sigma_3-\mu I \sigma_3
-W_{\bf k} \rho_2 I + \Delta_{\bf k} \rho_3 \sigma_1.
\ee
Here $\nu_m = 2 m \pi T$ and $\omega_n = (2n+1) \pi T$ are the bosonic
and fermionic Matsubara frequencies.
Notice that the cross-terms proportional to $t W_0$ cancel out each other.

After the frequency sum, the correlation function has the following form.
The first term, $\Lambda_{xx}^t$, of Eq.\ref{corr} which comes from 
the conventional current proportional to $t$ is obtained as follows
by taking $\nu \rightarrow 0$ first and ${\bf q} \rightarrow 0$ later.
\begin{eqnarray}
\Lambda^{t}_{xx}
&=& \sum_{\bf k} \frac{t^2 \sin{k_x}\sin{(k_x+q_x)}}{2}
\left[ \left(
1+\frac{A_1}{E_{1k} E_{1k+q}}-\frac{2 W_k^2}{\Omega_k^2}
\left(1+\frac{B_1}{E_{1k}E_{1k+q}} \right) \right)
\frac{f(E_{1k+q}) -f(E_{1k})}{E_{1k+q} - E_{1k}}
\right.
\cr
&&\hspace{2cm}  +
\left.
\left(
1-\frac{A_1}{E_{1k} E_{1k+q}}-\frac{2 W_k^2}{\Omega_k^2}
\left(1-\frac{B_1}{E_{1k}E_{1k+q}} \right) \right)
\frac{f(E_{1k+q}) +f(E_{1k}) -1}{E_{1k+q} + E_{1k}}
\right.
\cr
&&\hspace{2cm}  +
\left.
\left(
1+\frac{A_2}{E_{2k} E_{2k+q}}+\frac{2 W_k^2}{\Omega_k^2}
\left(1+\frac{B_2}{E_{2k}E_{2k+q}} \right) \right)
\frac{f(E_{2k+q}) - f(E_{2k}) }{E_{2k+q} - E_{2k}}
\right.
\cr
&&\hspace{2cm}  +
\left.
\left(
1-\frac{A_2}{E_{2k} E_{2k+q}}+\frac{2 W_k^2}{\Omega_k^2}
\left(1-\frac{B_2}{E_{2k}E_{2k+q}} \right) \right)
\frac{f(E_{2k+q}) +f(E_{2k}) -1}{E_{2k+q} + E_{2k}}
\right],
\label{corr_t}
\end{eqnarray}
\end{widetext}
where $f(x) = 1/(e^{x/T}+1)$ is the Fermi function and  
\begin{eqnarray}
A_{1,2} &=& \Delta_k^2 + \epsilon_k^2 - W_k^2 + \mu^2 
\mp \frac{2 \mu \epsilon_k^2}{\Omega_k},
\cr
B_{1,2} &= & -\epsilon_k^2-W_k^2 \pm  \mu \Omega_k.
\end{eqnarray}

On the other hand, the second term, $\Lambda^{W}_{xx}$, of Eq.\ref{corr}
comes from the additional contribution to the current due to the existence of
the dDW order parameter. After taking $\nu \rightarrow 0$ first and 
${\bf q} \rightarrow 0$ later, we get
\begin{widetext}
\begin{eqnarray}
\Lambda^{W}_{xx}
&=& \sum_{\bf k} \frac{W_0^2 \sin{(k_x)}\sin{(k_x+q_x)}}{2}
\left[ \left(
1+\frac{A_1^{\prime}}{E_{1k} E_{1k+q}}+\frac{2 W_k^2}{\Omega_k^2}
\left(1+\frac{B_1}{E_{1k}E_{1k+q}} \right) \right)
\frac{f(E_{1k+q}) -f(E_{1k})}{E_{1k+q} - E_{1k}}
\right.
\cr
&&\hspace{2cm}  +
\left.
\left(
1-\frac{A_1^{\prime}}{E_{1k} E_{1k+q}}+\frac{2 W_k^2}{\Omega_k^2}
\left(1-\frac{B_1}{E_{1k}E_{1k+q}} \right) \right)
\frac{f(E_{1k+q}) +f(E_{1k}) -1}{E_{1k+q} + E_{1k}}
\right.
\cr
&&\hspace{2cm}  +
\left.
\left(
1+\frac{A_2^{\prime}}{E_{2k} E_{2k+q}}-\frac{2 W_k^2}{\Omega_k^2}
\left(1+\frac{B_2}{E_{2k}E_{2k+q}} \right) \right)
\frac{f(E_{2k+q}) - f(E_{2k}) }{E_{2k+q} - E_{2k}}
\right.
\cr
&&\hspace{2cm}  +
\left.
\left(
1-\frac{A_2^{\prime}}{E_{2k} E_{2k+q}}-\frac{2 W_k^2}{\Omega_k^2}
\left(1-\frac{B_2}{E_{2k}E_{2k+q}} \right) \right)
\frac{f(E_{2k+q}) +f(E_{2k}) -1}{E_{2k+q} + E_{2k}}
\right],
\end{eqnarray}
\end{widetext}
where
\begin{equation}
A_{1,2}^{\prime} = \Delta_k^2 - \epsilon_k^2 + W_k^2 - \mu^2 
\pm \frac{2 \mu \epsilon_k^2}{\Omega_k}.
\end{equation}
Therefore, the superfluid density can be obtained from
Eq.\ref{super}, where $\Lambda_{xx}=\Lambda_{xx}^t+
\Lambda_{xx}^W$.

\section{Value of the ``quasiparticle charge'' from superfluid density}
At low temperatures, the leading paramagnetic contribution comes from 
the nodal quasiparticles and we can obtain the low energy
quasiparticle dispersion by expanding  
$\epsilon_k$ near $\mu$, and $\Delta_k$, $W_k$ near the node
as follows. 
\bea 
\epsilon_k - \mu &=&  v_F p_+,
\cr
\Delta_k &=& v_{\Delta} p_-,
\cr
W_k &=& v_W p_-,
\eea
where $p_+ = (p_x + p_y)/\sqrt{2}$ and $p_- = (p_x - p_y)/\sqrt{2}$
with the momentum, {\bf p}, measured from ${\bf k}_F$. 
Here, $v_F = t a/\sqrt{2}$, $v_{\Delta} = \Delta_0 a/\sqrt{2}$,
and $v_W = W_0 a /\sqrt{2}$. 
When $\mu=0$ or $\mu < T$, the quasiparticle
dispersion can be well described by
\be
E^2_{1k,2k} = v_F^2 p_+^2 + (v^2_{\Delta} + v^2_W) p_-^2. 
\label{disp}
\ee
Notice that the corresponding density of states has the following form.
\be
N(E) = \frac{E}{2\pi v_F \sqrt{v_{\Delta}^2+v_W^2}}.
\ee
Therefore, the specific heat as well as the measured gap
velocity in ARPES will be given by the combination of
$v_W$ and $v_{\Delta}$ through $\sqrt{v_{\Delta}^2+v_W^2}$.  
Now it is clear that the leading linear temperature dependence of the 
superfluid density comes from the terms with $d f(E_{1k})/d (E_{1k})$ factor. 
The coefficient of the linear temperature 
dependence, $\alpha$, is obtained as
\be
\alpha =
\frac{{\rm \ln 2}}{2\pi}  \frac{v_F}{\sqrt{v_{\Delta}^2+v_W^2}}
\left(1-\frac{v_W^2}{ 2 (v_{\Delta}^2+v_W^2)} + {\cal O}
(\frac{W_0^2}{t^2}) \right).
\label{alpha}
\ee

Comparing Eq.\ref{alpha} and Eq.\ref{alpha1}, and replacing $v_{\Delta}$ by 
the {\it measured} gap velocity, $\sqrt{v^2_{\Delta} + v^2_W}$, we find
\be 
Z_e^2 =  
\left(1-\frac{v_W^2}{ 2 (v_{\Delta}^2+v_W^2)} \right),
\label{ze}
\ee
where we neglect ${\cal O}(W_0^2/t^2)$ terms.
From the above result, one can easily see that $1/2 \le Z^2_e \le 1$.
If $W_0 \gg \Delta_0$, then $Z_e^2 \rightarrow 1/2$, while in the opposite 
case $\Delta_0 \gg W_0$ or $W_0 \rightarrow 0$, we get $Z_e^2 \rightarrow 1$.
Therefore, $Z_e$ varies from $1/\sqrt{2} (= 0.71)$ to 1 which 
is curiously close to the values in cuprates obtained by 
Ioffe and Millis \cite{millis}.

In the discussion above, we assumed $\mu < T$. 
In the opposite limit, $\mu > T$, the low energy quasiparticle 
dispersion is better described by
\be
E_{1k}^2 =(v_F p_+)^2 + (v_{\Delta} p_-)^2
\ee
which does not have any information about $W_k$, the dDW order parameter.
This is due to the fact that the quasiparticles recognize the presence
of the dDW gap only when their energy scale becomes larger than the
chemical potential. Carrying out the explicit computation of the current-current 
correlation function, we verified that the coefficient of the linear temperature
dependence of the superfluid density is given by Eq.\ref{alpha1} with
$Z_e=1$.\cite{han} 
Therefore, there is a cross-over in the value of $Z_e$ to unity 
when $\mu > T$ while $Z_e \not= 1$ for $\mu < T$. At the half-filling ($\mu=0$),
$Z_e$ is in general not unity. At any rate, if $\mu$ is smaller than $T$ in the 
experimentally relevant temperature range, the value of $Z_e$ is given by Eq.\ref{ze}.  

\section{conclusion}

In the strong coupling regime, the coupling between the supercurrent and the 
quasiparticles can be strongly renormalized by quasiparticle interactions
and lead to the deviation of $Z_e$ from the BCS value ($Z_e=1$).
Here we show that the deviation from $Z_e=1$ can also occur in the weak coupling
regime when the superconducting order coexists with an additional order 
parameter with the same internal symmetry. 
We have shown that in the case of the coexisting dSC and dDW states, the 
``quasiparticle charge'' is given by
$Z_e = \left( 1-\frac{v_W^2}{2 (v_{\Delta}^2+v_W^2)} \right)^{1/2}$
when the chemical potential $\mu$ is zero or $\mu < T$,
where $v_W$ and $v_{\Delta}$ are the gap velocities
determined by the maximum gaps of the dDW and dSC order parameters.
Thus $Z_e$ varies from $1/\sqrt{2}$ to 1.
The deviation of $Z_e$ from the unity is
a good measure of the additional coexisting order parameter. 
Our result of $Z_e \not= 1$ is also applicable to the systems where a 
superconducting order parameter coexists with another order parameter with 
the same internal symmetry; e.g., $p$-wave superconductor coexisting with 
$p$-density wave, although the expression of $Z_e$ would be different
for different systems.

Our results suggest that the determination of
the value of $Z_e$ alone will not sharply distinguish two possibilities;
Mott physics and the existence of a hidden or an additional order parameter.
However, if there exists a superconductor far away from a Mott insulator
while it exhibits $Z_e < 1$, one would strongly suspect that there might
exist a hidden order parameter. The direct relevance of our results for the
coexisting dSC and dDW states to cuprates has to discussed in conjunction 
with other experimental data and it is beyond the scope of this paper.
 
{\it Acknowledgments:}
This work was supported in part by Canadian Institute for 
Advanced Research (H.Y.K. and Y.B.K.) and Alfred P. Sloan Foundation (Y.B.K.).


\begin{references}
\bibitem{relevance}
It was proposed that the d-density wave (dDW) may exist in 
underdoped cuprates as a ``hidden order parameter'' and may be responsible 
for the pseudogap phenomena. \cite{sudip00,hykee,marston88}
\bibitem{han} Q. H. Wang, J. H. Han, and D. H. Lee, Phys. Rev. Lett. {\bf 87},
077004(2001); They discussed the behavior of $\alpha$ and concluded that
it has the same form as that of dSC. We showed that $\alpha$ has a different
form when $\mu=0$ or $\mu < T$. We believe that their conclusion of diverging 
$\alpha$ as $\Delta \rightarrow 0$ is not correct because the upper cutoff
of the Dirac dispersion is given by $\Delta$, thus the linear temperature
dependence occurs only when $T < \Delta$. As $\Delta$ goes to zero, the linear 
temperature dependence of the superfluid density simply disappears.
\end{references}
\end{document}